\documentclass[aps,prl,twocolumn,floats,showpacs]{revtex4}
\usepackage{bbm}
\usepackage{amssymb}
\usepackage{ifpdf}
\usepackage{graphicx,epsfig}
\newcommand{\Z}{{\mathbb{Z}}}

\begin{document}
\title{From Decay to Complete Breaking: Pulling the Strings in $SU(2)$ 
Yang-Mills Theory}
\author{M.\ Pepe$^a$ and U.-J.\ Wiese$^{b,c}$}
\affiliation{
$^a$ INFN, Istituto Nazionale di Fisica Nucleare, Sezione di Milano-Bicocca \\
Edificio U2, Piazza della Scienza 3 - 20126 Milano, Italy \\
$^b$ Center for Research and Education in Fundamental Physics,
Institute for Theoretical Physics, Bern University,
Sidlerstr.\ 5, 3012 Bern, Switzerland \\
$^c$ Institute for Theoretical Physics, ETH Z\"urich,
Schafmattstr.\ 32, CH-8093 Z\"urich, Switzerland}


\begin{abstract}

We study $\{2Q+1\}$-strings connecting two static charges $Q$ in $(2+1)$-d 
$SU(2)$ Yang-Mills theory. While the fundamental $\{2\}$-string between two 
charges $Q = \frac{1}{2}$ is unbreakable, the adjoint $\{3\}$-string connecting
two charges $Q = 1$ can break. When a $\{4\}$-string is stretched beyond a 
critical length, it decays into a $\{2\}$-string by gluon pair creation. When a
$\{5\}$-string is stretched, it first decays into a $\{3\}$-string, which 
eventually breaks completely. The energy of the screened charges at the ends of
a string is well described by a phenomenological constituent gluon model.

\end{abstract} 


\maketitle

Studies of the strings connecting two static color charges provide valuable 
insights into the physics of confinement in $SU(N)$ Yang-Mills theories.
The properties of the string connecting a static quark-anti-quark pair with 
charges in the fundamental $\{N\}$ and $\{\overline{N}\}$ representations are 
described by a low-energy effective bosonic string theory. While the string 
tension $\sigma$ determines the quark-anti-quark potential $V(r) = \sigma r$ at
asymptotic distances, the massless modes corresponding to transverse 
fluctuations of the string give rise to the universal L\"uscher term 
proportional to $1/r$ \cite{Lue80,Lue81}, as well as to a diverging string
thickness proportional to $\log r$ \cite{Lue81a}. The effective string theory 
also makes detailed predictions for the excited states of the string 
\cite{Lue04b}. Lattice gauge theory provides us with a powerful tool for 
investigating the string dynamics using Monte Carlo simulations. In this way, 
the linearly rising quark-anti-quark potential has been calculated at large
distances \cite{Bal92}. By developing a highly efficient multi-level 
simulation technique \cite{Lue04}, L\"uscher and Weisz have studied the 
universal $1/r$-term in the quark-anti-quark potential at large distances 
\cite{Lue04a}. 

In theories with dynamical fundamental charges, the confining string connecting two static
color charges can break due to the creation of dynamical charge-anti-charge pairs which
screen the external static sources. Numerical evidence for charge screening was obtained
in a lattice gauge-Higgs model in which the dynamical fundamental charges are scalars
\cite{Boc90}. Direct evidence for string breaking was first observed in the $SU(2)$
gauge-Higgs model \cite{Kne98,Phi99} and later also in the $\Z(2)$ gauge-Higgs model
\cite{Gli05}. While the fundamental string is unbreakable in $SU(N)$ Yang-Mills theory,
the string connecting static adjoint charges can break due to pair-creation of dynamical
gluons. This effect has been investigated in
\cite{Pou97,Ste99,Phi99a,Kra03,For00,Kal02,Gre07}. Numerical evidence for string breaking
in lattice QCD with dynamical quarks has been observed in \cite{Bal05}. The center
symmetry of $SU(N)$ Yang-Mills theory is $\Z(N)$.  Consequently, each $SU(N)$
representation and hence each external static charge can be characterized by its $N$-ality
$k = \{0,1,...,N-1\}$.  Strings connecting external charges with $N$-ality $k \neq 0$ are
known as $k$-strings. These strings are unbreakable and have a $k$-dependent string
tension, which may or may not be proportional to the Casimir operator of the corresponding
representation \cite{Ole81,Amb84}.

Since it is easiest to simulate numerically, in this letter we study 
the dynamics of strings in $(2+1)$-d $SU(2)$ Yang-Mills theory which has the 
center $\Z(2)$. Other theories in $(3+1)$ dimensions or with other gauge groups
are expected to show similar behavior. Here we investigate the strings 
connecting two static charges $Q$ in the $SU(2)$ representation $\{2Q+1\}$, 
which we refer to as $\{2Q+1\}$-strings, not to be confused with $k$-strings. 
The $\{2Q+1\}$-strings with integer $Q$ have $k=0$ and will eventually break at
large distances, while the $\{2Q+1\}$-strings with
half-integer $Q$ have $k = 1$ and are unbreakable. At asymptotic distances all 
$\{2Q+1\}$-strings connecting half-integer charges have the same tension 
$\sigma$ as the fundamental $\{2\}$-string. Since $SU(2)$ Yang-Mills theory
has no dynamical fundamental charges, the static charges $Q$ at the two ends of
a $\{2Q+1\}$-string can only be screened by dynamical gluons. When a pair of 
gluons is created from the vacuum, the external sources are screened and thus 
reduced to $Q-1$. As a consequence, the $\{2Q+1\}$-string decays to a
$\{2Q-1\}$-string and abruptly reduces its tension accordingly 
\cite{Arm03,Gli05a}. While some numerical evidence for string decay was 
presented in \cite{Del03}, using the multi-level simulation technique of 
\cite{Lue04}, we are able, for the first time, to investigate string decay in 
detail. 

We consider $(2+1)$-d $SU(2)$ Yang-Mills theory on a cubic lattice using the
standard Wilson plaquette action for link variables in the fundamental
representation. The external color charges $Q$ at the two ends of a string are 
represented by Polyakov loops $\Phi_Q(x)$ in the $\{2Q+1\}$ representation 
wrapping around the Euclidean time direction. The corresponding potential 
$V_Q(r)$ between the static sources is extracted from the Polyakov loop 
correlator 
\begin{equation}
\langle \Phi_Q(0) \Phi_Q(r) \rangle \sim \exp(- \beta V_Q(r)).
\end{equation}
In order to ensure a good projection on the ground state of the string, we have
simulated at inverse temperatures as large as $\beta = 64$ in lattice units. 
The spatial lattice size was $L = 32$ and the bare gauge coupling was 
chosen as $4/g^2 = 6.0$ which puts the deconfinement phase transition at 
$\beta_c \approx 4$. While this is a moderate coupling, we are confident that 
our results remain unchanged, at least qualitatively, in the continuum limit. 
The values of the simulated Polyakov loop correlators range from $10^{-8}$ to
$10^{-135}$. Measuring such small signals would be completely impossible 
without the L\"uscher-Weisz multi-level simulation technique. We have slightly 
refined this method by applying the segmentation of the lattice not only to 
slabs in time, but also to blocks in space. By carefully tuning the parameters 
of the multi-level algorithm, we have been able to extract the potentials 
$V_Q(r)$ for the $\{2\}$-, $\{3\}$-, $\{4\}$-, and $\{5\}$-strings. As shown in 
the top panel of figure 1, at distance $r \approx 8$, the $\{4\}$-string 
decays, thus reducing its tension to the one of the fundamental $\{2\}$-string.
Similarly, the bottom panel shows that, at distance $r \approx 6$, the 
$\{5\}$-string decays and reduces its tension to the one of the adjoint 
$\{3\}$-string. Only at $r \approx 10$ the string breaks completely, at about 
the same distance as the adjoint $\{3\}$-string. Not unexpectedly, the tension 
of a string is the same, no matter whether it connects screened or unscreened 
external charges $Q$.
 
\begin{figure}[t]
\includegraphics[width=0.44\textwidth]{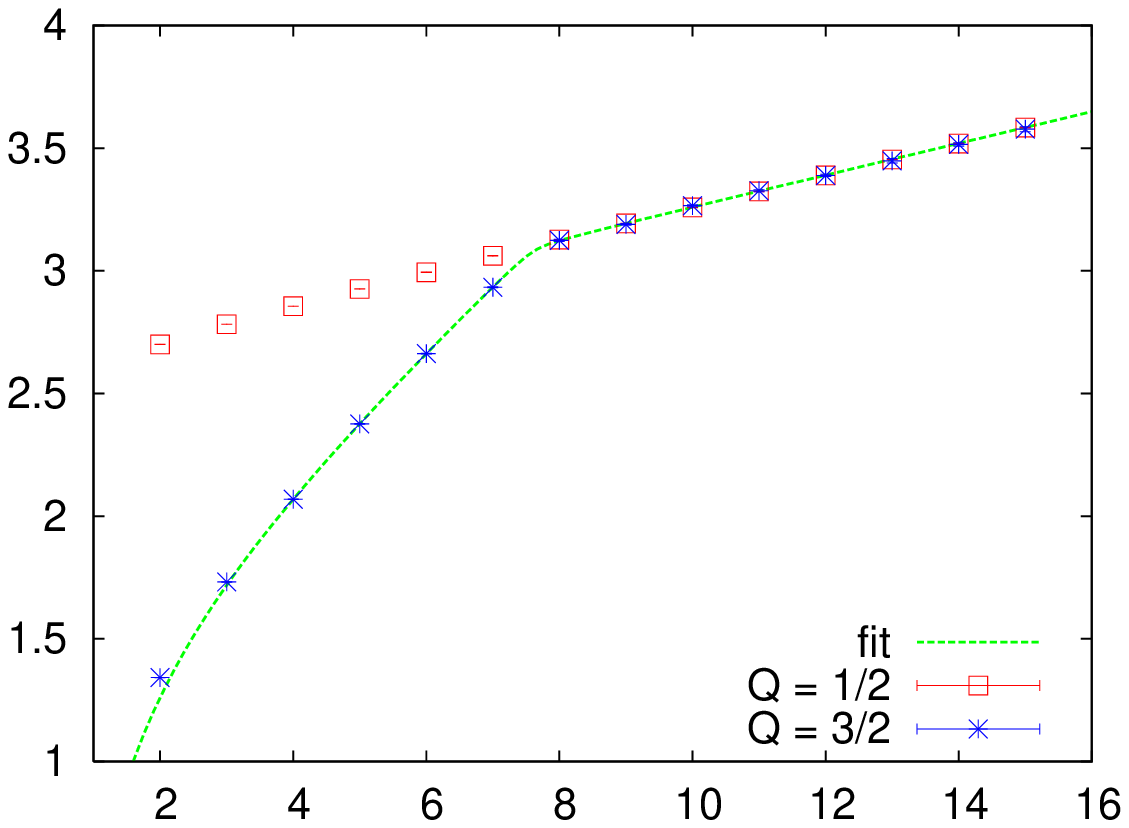}
\vskip-.3cm
\includegraphics[width=0.44\textwidth]{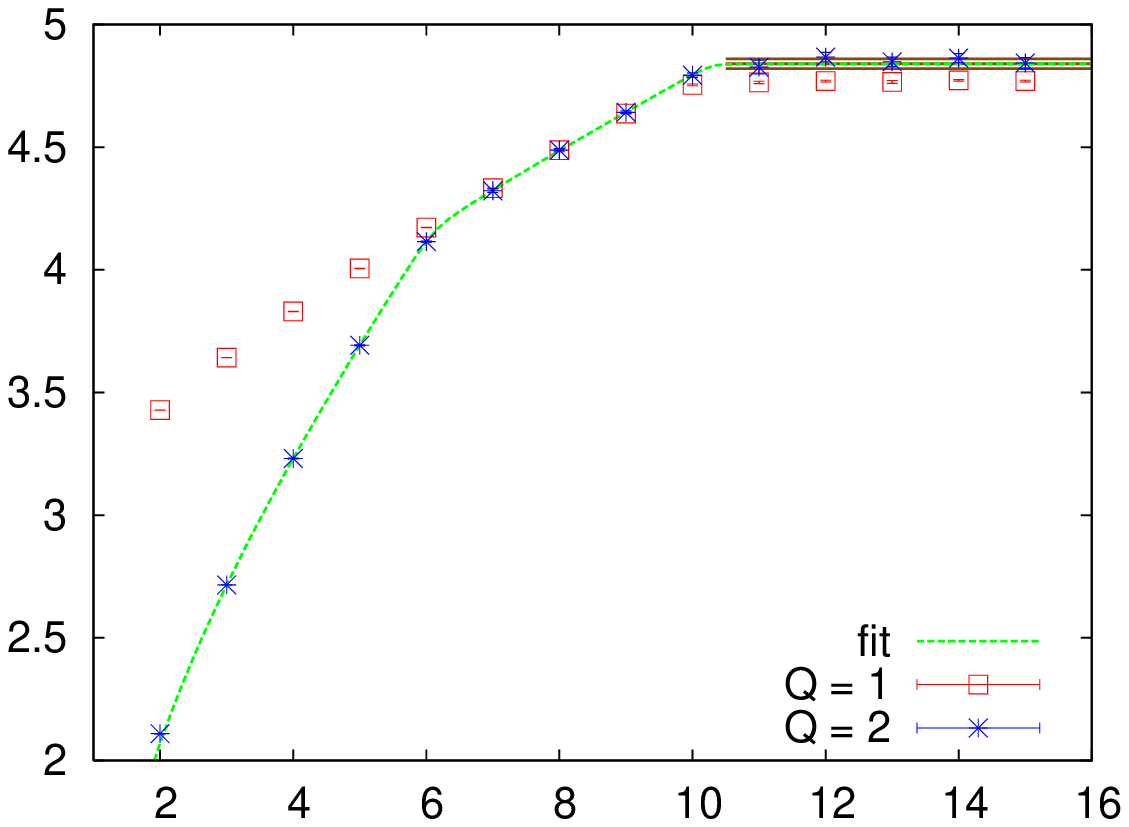}
\vskip-.3cm
\caption{\it Top: Potential $V(r)$ of two static color charges with 
$Q = \frac{1}{2}$ (squares) and $Q = \frac{3}{2}$ (stars), shifted by a 
constant for a more convenient comparison of the slopes. Bottom: The same for
$Q = 1$ (squares) and $Q = 2$ (stars). The lines are a fit of the multi-channel
model to the Monte Carlo data. The horizontal band at $2 M_{0,2}=4.84(2)$ corresponds to
twice the mass of a source of charge $Q=2$ obtained from the measurement of a single
Polyakov loop.}
\end{figure}

A fit of the fundamental potential to
\begin{equation}
V_{1/2}(r) = \sigma r - \frac{\pi}{24 r} + 2 M + {\cal O}(1/r^3),
\end{equation}
works very well and yields the asymptotic string tension $\sigma = 0.06397(3)$.
In particular, the Monte Carlo data are in excellent agreement with the 
predicted coefficient $- \frac{\pi}{24}$ of the L\"uscher term. The ``mass'' 
contribution of an external charge $Q = \frac{1}{2}$ to the total energy of the
system is given by $M = 0.109(1)$. This ``mass'' itself is not physical because
it contains ultra-violet divergent pieces. Since string decay occurs at 
moderate distances, its typical energy scale is not well separated from 
$\Lambda_{\text{QCD}}$. Consequently, unlike the string behavior at asymptotic 
distances, string decay can not be addressed in a fully systematic low-energy 
effective string theory. In particular, unlike the string tension $\sigma$ of 
the unbreakable fundamental string, the tension $\sigma_Q$ of an ultimately 
unstable $\{2Q+1\}$-string (with $Q \geq 1$) is not defined unambiguously. Here
we define $\sigma_Q$ by a fit of the Monte Carlo data to a simple
phenomenological model. In this model, we consider the $\{2Q+1\}$-string as a 
multi-channel system. A channel containing a $\{2Q+1\}$-string connecting two 
charges $Q$, which resulted from screening a larger charge $Q+n$ by $n$ gluons,
has the energy
\begin{equation}
E_{Q,n}(r) = \sigma_Q r - \frac{c_Q}{r} + 2 M_{Q,n}.
\end{equation}
Here $c_Q$ is the coefficient of a sub-leading $1/r$ correction which does not
necessarily assume the asymptotic L\"uscher value $- \frac{\pi}{24}$. The 
``mass'' $M_{Q,n}$ describes the contribution of an original charge $Q+n$ that 
has been screened to the value $Q$ by $n$ gluons. Just as the ``mass'' 
$M = M_{1/2,0}$, the ``masses'' $M_{Q,n}$ themselves are not physical, because 
they again contain ultra-violet divergent contributions. However, the mass 
differences $\Delta_{Q,n} = M_{Q-1,n+1} - M_{Q,n}$ are physical since the 
divergent pieces then cancel. The $\{3\}$- and $\{4\}$-string are described by 
the two-channel Hamiltonians $H_1$ and $H_{3/2}$, while the $\{5\}$-string is
described by the three-channel Hamiltonian $H_2$ with 
\begin{eqnarray}
&&H_1(r) = \left(\begin{array}{cc} E_{1,0}(r) & A \\ A & E_{0,1}(r)
\end{array}\right), \nonumber \\
&&H_{3/2}(r) = \left(\begin{array}{cc} E_{3/2,0}(r) & B \\ B & E_{1/2,1}(r)
\end{array}\right), \nonumber \\
&&H_2(r) = \left(\begin{array}{ccc} E_{2,0}(r) & C & 0 \\ C & E_{1,1}(r) & A \\
0 & A & E_{0,2}(r) \end{array}\right).
\end{eqnarray}
Here $A$, $B$, and $C$ are decay amplitudes which we assume to be 
$r$-independent.
The potential $V_Q(r)$ is the energy of the ground state of $H_Q$. Figure 2 
compares the forces $F(r) = - dV(r)/dr$ in the $\{2\}$-, $\{3\}$-, $\{4\}$-, 
and $\{5\}$-string cases with the results of the multi-channel model. The 
tensions $\sigma_Q$ listed in table 1 have been determined by a fit to the 
Monte Carlo data. The simple model works rather well. It is interesting to note
that the ratios $\sigma_Q/\sigma$ do not obey Casimir scaling, i.e.\ they are 
not equal to $4 Q(Q+1)/3$. The ``masses'' $M_{Q,n}$ are listed in table 2.
Remarkably, within the error bars, the mass differences $\Delta_{Q,0} = 
M_{Q-1,1} - M_{Q,0}$ all take the same value $M_G = 0.65(5)$, independent of 
$Q$. We interpret $M_G$ as a constituent gluon mass which in units of the 
string tension takes the value $M_G/\sqrt{\sigma} = 2.6(2)$. It should be 
pointed out that, in contrast to the string tension, $M_G$ is not unambiguously
defined. It just results from the fit parameters of the phenomenological model.
The value $\Delta_{1,1} = M_{0,2} - M_{1,1} = 0.71(3)$ indicates that the
addition of a second constituent gluon costs an energy slightly larger than 
$M_G$. Interestingly, the mass of two constituent gluons $2 M_G = 1.3(1)$ is 
close to the $0^+$ glueball mass $M_{0^+} = 1.198(25)$ obtained in \cite{Mey03}
at the same value of the bare coupling. $M_G$ also sets the distance scale for
string decay and string breaking. A leading order estimate for the distance at which the
$\{4\}$-string decays into the $\{2\}$-string is $r \approx 2 M_G/(\sigma_{3/2} -
\sigma_{1/2}) = 7.3(6)$, while the distance at which the $\{3\}$- and the $\{5\}$-string
ultimately break is estimated to be around $r \approx 2 M_G/\sigma_1 = 9.0(7)$.

\begin{figure}[t]
\includegraphics[width=0.44\textwidth]{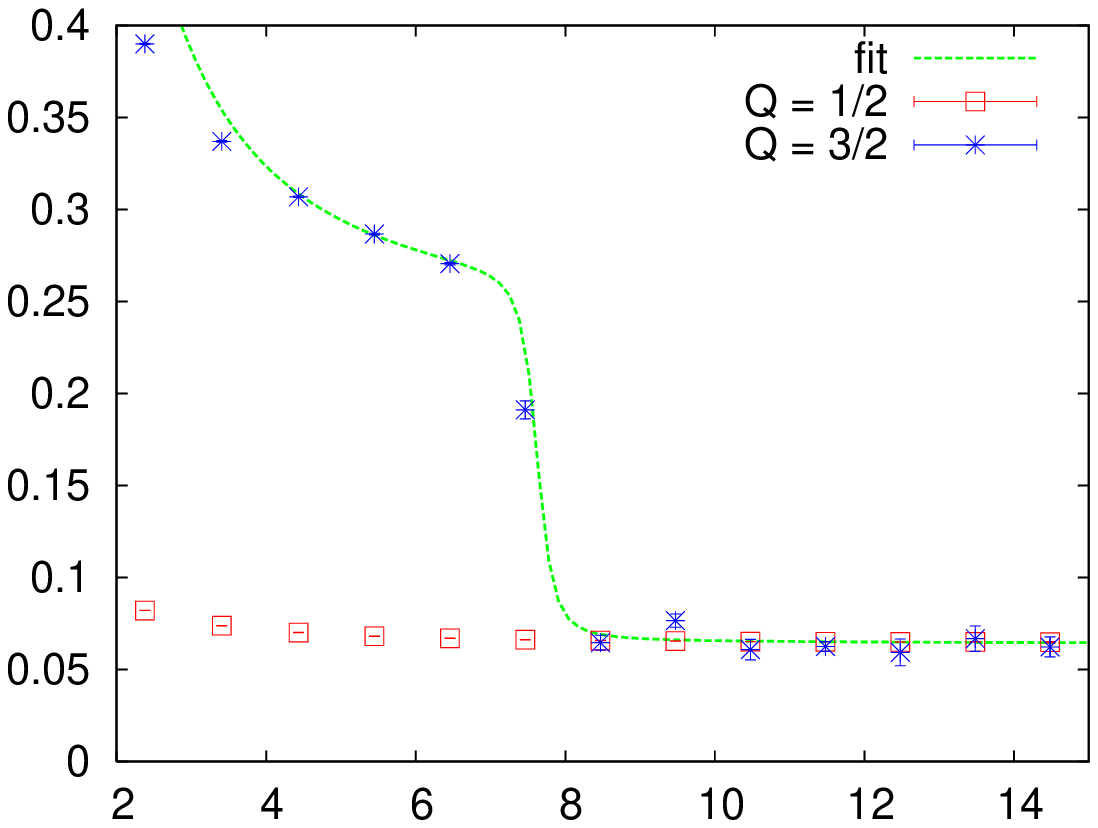}
\vskip-.3cm
\includegraphics[width=0.44\textwidth]{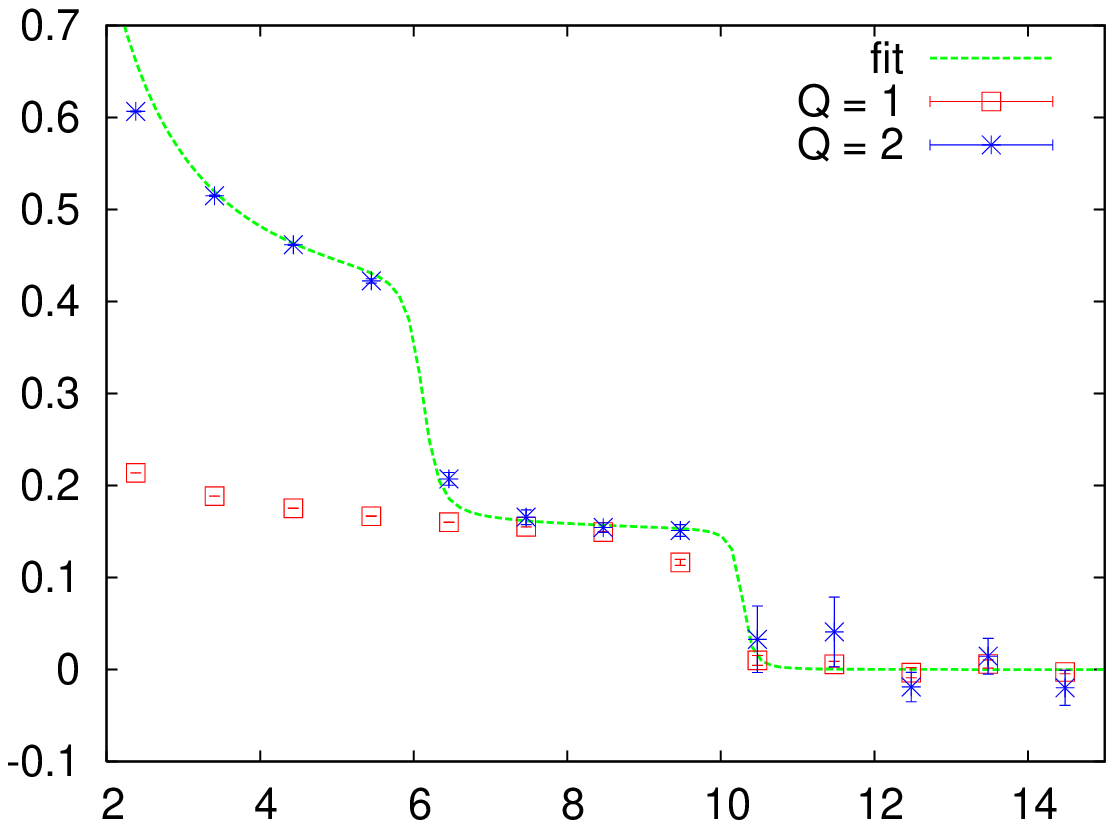}
\vskip-.3cm
\caption{\it Top: Forces $F(r)$ that the $\{2\}$- and $\{4\}$-string exert on 
the external charges $Q = \frac{1}{2}$ (squares) and $Q = \frac{3}{2}$ (stars),
respectively. Bottom: The same for the $\{3\}$- and $\{5\}$-string connecting 
external charges $Q = 1$ (squares) and $Q = 2$ (stars), respectively. The lines
represent the fit of the multi-channel model to the Monte Carlo data.}
\end{figure}

\begin{table}
\begin{center}
\begin{tabular}{|c|c|c|c|}
\hline
$Q$ & $\sigma_Q$ & $\sigma_Q/\sigma$ & $4 Q(Q+1)/3$ \\
\hline
\hline
1/2 & 0.06397(3) & 1       & 1   \\ 
\hline
1   & 0.144(1)   & 2.25(2) & 8/3 \\ 
\hline
3/2 & 0.241(5)   & 3.77(8) & 5   \\ 
\hline
2   & 0.385(5)   & 6.02(8) & 8   \\ 
\hline
\end{tabular}
\end{center}
\caption{\it Fitted values of the string tensions $\sigma_Q$. The ratio 
$\sigma_Q/\sigma$ with $\sigma = \sigma_{1/2}$  is compared with the value 
$4 Q(Q+1)/3$ representing Casimir scaling.}
\end{table}

\begin{table}
\begin{center}
\begin{tabular}{|c|c|c|c|c|c|}
\hline
$Q$ & $M_{Q,0}$ & $M_{Q-1,1}$ & $M_{Q-2,2}$ & $\Delta_{Q,0}$ & $\Delta_{Q-1,1}$ \\
\hline
\hline
1/2 & 0.109(1) & ---      & ---     & ---     & ---     \\ 
\hline
1   & 0.37(3)  & 1.038(1) & ---     & 0.67(3) & ---     \\ 
\hline
3/2 & 0.72(5)  & 1.32(5)  & ---     & 0.60(5) & ---     \\ 
\hline
2   & 1.04(3)  & 1.71(3)  & 2.42(1) & 0.67(3) & 0.71(3) \\ 
\hline
\end{tabular}
\end{center}
\caption{\it Fitted values of the ``mass'' $M_{Q,n}$ of an original charge 
$Q+n$ that has been screened to the value $Q$ by $n$ gluons, together with the
differences $\Delta_{Q,n} = M_{Q-1,n+1} - M_{Q,n}$.}
\end{table}

String decay can be viewed as a quantum analogue of the classical process of strand
rupture in a cable consisting of a bundle of strands. When such a cable is stretched
further and further, individual strands eventually rupture, thereby abruptly reducing the
tension of the cable. While strand rupture is well-known in the material science of
centimeter thick steel cables with a tension of about $10^5$ Newton, we have seen that a
similar process occurs for the confining strings in non-Abelian gauge theories which have
about the same tension but are 13 orders of magnitude thinner. Whether a strand picture
may correctly describe the actual anatomy of decaying $\{2Q+1\}$-strings is an interesting
question that will require further investigations which go beyond the scope of the present
letter.

It would be interesting to investigate string decay and string breaking
for other $SU(N)$ gauge theories. In $SU(3)$ Yang-Mills theory the
$\{3\}$-string connecting a quark in the $\{3\}$ with an anti-quark in the 
$\{\overline 3\}$ representation is unbreakable, while the $\{8\}$-string 
connecting two adjoint sources can break by pair creation of gluons. When a
$\{6\}$-string is stretched, the external source in the $\{6\}$-representation 
will eventually be screened to a $\{\overline 3\}$ by a gluon. The 
corresponding string decay should be analogous to the decay of the 
$\{4\}$-string in $SU(2)$ Yang-Mills theory discussed above. In analogy to the 
$\{5\}$-string in $SU(2)$, the $\{10\}$-string in $SU(3)$ Yang-Mills theory is 
expected to decay to an adjoint $\{8\}$-string, before it breaks completely at 
larger distances. In QCD with dynamical quarks, strings can also decay by 
quark-anti-quark pair creation. Due to its $\Z(4)$ center symmetry, $SU(4)$ 
Yang-Mills theory has two distinct unbreakable strings, connecting external 
charges either in the $\{4\}$ and $\{\overline 4\}$ or in the 
$\{6\}$-representation. For external sources with non-trivial $N$-ality, one 
then expects cascades of string decays down to the $\{4\}$-string for $k = 1,3$
and down to the $\{6\}$-string for $k = 2$. 

Studying gauge groups other than $SU(N)$ would also be interesting. For
example, all $Sp(N)$ gauge theories have the same center $\Z(2)$. The first Lie
group in this sequence is $Sp(1) = SU(2) = Spin(3)$, while the second is
$Sp(2) = Spin(5)$, the universal covering group of $SO(5)$. In $Sp(2)$ 
Yang-Mills theory, only the fundamental $\{4\}$-string is absolutely stable. As
usual, the adjoint $\{10\}$-string can break by pair creation of gluons. The 
representation $\{5\}$ is center-neutral. Since in $Sp(2)$
\begin{equation}
\{5\} \otimes \{10\} = \{5\} \oplus \{10\} \oplus \{35\},
\end{equation}
a single gluon can screen a charge $\{5\}$ only to a $\{10\}$ or a $\{35\}$. We
expect the unstable $\{5\}$-string to have a smaller tension than the adjoint 
$\{10\}$-string or the $\{35\}$-string. In that case, the $\{5\}$-string will 
break in one step by the creation of four gluons, without any intermediate 
string decay.

Finally, it would be interesting to investigate the importance of the center
for the phenomenon of string decay. The exceptional group $G(2)$ is the 
simplest Lie group with a trivial center. Still, $G(2)$ Yang-Mills theory 
confines color (although without an asymptotic string tension) \cite{Hol03}. 
Furthermore, it has a first order deconfinement phase transition 
\cite{Pep05,Pep07}. 
In fact, as we have discussed in the context of $Sp(N)$ Yang-Mills theories, 
the order of the deconfinement phase transition is controlled by the size of the 
gauge group and not by the center \cite{Hol04}. In $G(2)$ Yang-Mills theory, 
even a charge in the fundamental $\{7\}$ representation can be screened by 
gluons in the adjoint $\{14\}$ representation. As a result, there are no 
unbreakable strings. Since in $G(2)$
\begin{equation}
\{7\} \otimes \{14\} = \{7\} \oplus \{27\} \oplus \{64\},
\end{equation}
a single gluon can screen a charge $\{7\}$ only to a $\{27\}$ or a $\{64\}$. In
$G(2)$ Yang-Mills theory, approximate Casimir scaling has been verified for 
unstable strings including the $\{27\}$- and the $\{64\}$-string \cite{Lip08}. 
As a consequence, the fundamental $\{7\}$-string is stable against decay due to
the creation of a single pair of gluons. The same is true even for processes 
involving four gluons. Based on the group theory of $G(2)$, we expect the 
fundamental $\{7\}$-string to break due to the simultaneous creation of six 
gluons, without any intermediate string decay.

Using the L\"uscher-Weisz multi-level algorithm, studying string decay in 
$SU(3)$, $SU(4)$, $Sp(2)$, $G(2)$, and other Yang-Mills theories is interesting
and definitely feasible. One may also ask whether string decay can be studied 
analytically in supersymmetric theories.

M.P. acknowledges useful discussions with F.~Gliozzi and J.~Greensite.
This work is supported in part by funds provided by the Schweizerischer 
Nationalfonds (SNF). The ``Center for Research and Education in Fundamental 
Physics'' at Bern University is supported by the ``Innovations- und 
Kooperationsprojekt C-13'' of the Schweizerische 
Uni\-ver\-si\-t\"ats\-kon\-fe\-renz (SUK/CRUS).

\end{document}